\documentclass[12pt]{article}
\usepackage{amssymb,amsfonts}
\newcommand{\la}{\lambda}
\newcommand{\sx}{\sigma_1}
\newcommand{\sy}{\sigma_2}
\newcommand{\sz}{\sigma_3}

\begin{document}
\vskip0.3cm
\centerline{\bf {\large {{The Inverse Scattering Transform for a Model }}}}
\centerline{{\bf{\large of Colomb's plasma with the negative temperature}}}
\vskip0.3cm
\centerline{\bf {E. Sh. Gutshabash}}
\vskip0.6cm

\hfil{Institute Research for Physics, Sankt-Petersburg State University, Russia}\hfil                 

\hfil{e-mail: gutshab@EG2097.spb.edu}\hfil
\vskip0.5cm
\centerline{\bf Abstract}
\vskip0.4cm
\hskip 2cm \parbox {13cm}{\small The boundary problem for a two-dimensional
elliptical equation -$\sinh$-Gordon has been investigated. The exact
solutions have been found and identities of traces have been
proposed. The application of the problem to the model of the
Coulomb's plasma with the negative temperature has been considered.}

\vskip0.8cm

\centerline{\bf {1. Introduction}}
\vskip0.6cm
This paper is devoted to research and the construction of the
explicit solutions of the nonlinear elliptical equation
-$\sinh$-Gordon:

$$
\triangle u=-4\sinh u ,\eqno(1.1)
$$
where $\triangle$ is two-dimensional Laplas's operator.

On the one hand, it is of a great mathematical interest due to is
describes the immersion of a negative curvature surface in the
three-dimensional Euclidean space [1], and on the other hand it has a
number of physical applications. In particular, it arises as the
model of two-dimensional Colomb's system at the negative temperatures
[2,3].

L.Onzager appears to have been first who introduced a conception of
negative temperatures in the context of a problem of a vortical line
description. Further the conception turned out to be connected with
so-called anomalous systems consisting of weakly interacting
particles, spins and so on (see, for example [4]) where an energetic
spectrum is bounded above and some maximal value of energy is
available. In this case a statistical sum of the system is $\sum_i
g(E_i)\exp(-E_i/T)$ under assumption that $g(E_i) \sim
\exp(-aE_i),\; E_i
\to \infty,\; a > 0$, where $g(E_i)$ is a density of a statements number
, and $E_i$ is an energetic level that can be finite both at $T
> 0$ and $ T< 0$ ($T$ is absolute temperature). Assuming
that $a=\infty$ and hence $g(E_i)=0$ at $E_i > E_{\max}$, it is not
difficult to see that the statistical sum changes into a finite
series and has a finite value at $T \in (-\infty, \infty)$. Then it
is obvious that the density of the energetic levels that is
proportional to  Boltsmann's factor $\exp(-E_i/T)$, at $ T >0$ drops
as the energy increases.

In particular, at $T \to
\infty \: \exp(-E_i/T) \to 1$, i.e. all levels proved to be distributed
uniformly. Another picture is at $T \in (-\infty, 0)$, in this case
Boltsmann's factor is an increasing function of $E_i$ and there
exists an phenomenon of an inversion of a levels distribution with
the density at the up levels becoming more than at down ones. It is
clear, that the equation (1.1) involving the difference of
Boltsmann's factors for the one-charged particles in its right side
exactly corresponds to the situation of a negative temperatures that
was described above.

We now turn to a statement the boundary problem. Let us suppose that
the function $u(x,y)$  determined on the half-plane $\mathbb
R_{+}^{2}=
\{(x,y): x \in {(-\infty,\infty)},\: y \ge 0\}$, is the real-valued and
enough smooth, $u(x,0),\:u_y(x,0)$ are boundary values of the desired
initial function and its normal derivative correspondently. Also we
assume that one of these condition is known, but there is some
relation of the condition to one another that will by some given
below (a nonlinear analog of the condition of the third kind).

Moreover, we will assume, that

$$
u(x,0) \to 0 \:\:\:at \:\:|x| \to \infty. \eqno(1.2)
$$

Equation (1.1) is integrable  and can be represented in the form of
the compatibility condition of the linear matrix system:

$$
\Psi_x(x,y,\lambda)=U(x,y,\lambda)\Psi(x,y,\lambda),\eqno(1.3a)
$$

$$
\Psi_y(x,y,\lambda)=V(x,y,\lambda)\Psi(x,y,\lambda).\eqno(1.3b)
$$
Here $\Psi(x,y,\lambda)$ is a matrix-valued ($2\times 2$) function,
$\la \in
\mathbb C$
is a spectral parameter, $U$ and $V$ will look like:

$$
U(\la)\equiv U(x,y,\la)=i[\frac {2}{\la^2-1}+\frac {\la+1}{2(\la-1)}
(\cosh u-1)]\sz+\frac {\la+1}{2(\la-1)}\sinh ue^{-2ix\sz}\sy+
$$
$$
+\frac {u_z}{2}e^{-2ix\sz}\sx ,
$$
$$
V(\la)\equiv V(x,y,\la)=[\frac {2\la}{\la^2-1}+i\frac {\la+1}{2(\la-1)}
(\cosh u-1)]\sz+\frac {i(\la+1)}{2(\la-1)}\sinh ue^{-2ix\sz}\sy+
$$
$$
+\frac {iu_z}{2}e^{-2ix\sz}\sx .
$$
It should be notice that a rather nontrivial and complex form of the
gauge of the operators $U$ and $V$ is connected with the gauge
equivalence of the equation (1.1) and two-dimensional isotropic
Heisenberg's ferromagnet, which was proved earlier [5,6]. The choice
$\pm 1$ as the poles is necessary to obtain a convenient system of
contours in solving Riemann's problem (see below).

\vskip0.6cm
\centerline{{\bf {2. Associated linear problem}}}
\vskip0.6cm
Let us $\Psi^{\pm}(x,\la)$ are the solutions of the (1.3) with the
conditions ($k(\la)=2/(\la^2-1)$):

$$
\Psi^{\pm}(x,\la)\to \exp(ik(\la)x\sz)(1+o(1)), \;\; x\to \pm \infty .\eqno(2.1)
$$
Let us determine a transition matrix $T(\la) \equiv T(\la,y)\;
(\la^2=\bar
\la^2)$:
$$
\Psi^{-}(x,\la)=\Psi^{+}(x,\la)T(\la),\:\: T(\la)=\left (\begin{array}{cc}
                               a(\la)&b(\la)\\
                               c(\la)&d(\la)
                        \end{array}\right) .\eqno(2.2)
$$
The matrix $T(\la)$ is unimodular: $\det T(\la)=1$.

Assuming $\phi^{\pm}(x,\la)=
\Psi^{\pm}(x,\la)\exp(-ik(\la)x\sz)$, from (1.3a) we obtain:
$$
\phi^{\pm}(\la)=e^{-ix\sz}e^{(u/2)\sx}\sy e^{-ix\sz}\bar \phi^{\pm}
(-\bar \la)\sy ,\
\la^2=\bar \la^2,      \eqno(2.3)
$$
$$
\phi^{\pm}(\la)=e^{-2ix\sz}\sx\phi^{\pm}(\frac {1}{\la})\sx ,\;\;
\la^2=\bar \la^2 . \eqno(2.4)
$$
From these relations it follows the relations of symmetry

$$
T(\la)=\sy\bar T(-\bar \la)\sy, \;\; \la^2=\bar \la^2, \eqno(2.5)
$$
$$
T(\la)=\sx\bar T(\frac {1}{\la})\sx . \eqno(2.6)
$$
Then we have

$$
T(\la)=\sz\bar T(-\frac {1}{\bar \la})\sz,  \;\;
\la^2=\bar \la^2 .\eqno(2.7)
$$

We find a dependence of scattering data from the variable $y$. For
this purpose let us turn to the system (1.3b). On fulfilling standard
operations and taking $l(\la)=2\la/(\la^2-1)=\la k(\la)$, we obtain:

$$
a(\la,0)=a(\la,y), \; \; \;  b(\la,y)=b(\la,0)e^{2l(\la)y} ,
\eqno(2.8)
$$
$$
c(\la,y)=c(\la,0)e^{-2l(\la)y},  \; \; \;    d(\la,0)=d(\la,y).
\eqno(2.9)
$$
It follows from these relations that the coefficients $a(\la)$ and
$d(\la)$ act as the produced functional of "the integrals of
movement". Alternatively, from the requirement of finiteness of the
coefficients $b(\la)$ and $c(\la)$ it follows, that it is necessary
to put $b(\la,0)$ equal to zero at

$$
\la > 1,  \; \; \; -1<\la<0, \eqno(2.10)
$$
and put $c(\la,0)$ equal to zero at

$$
0< \la <1,  \;  \; \;   \la<-1. \eqno(2.11)
$$
These inequalities determine boundaries of the zones. Domains in the
real axis that are additional to these zones are continuous spectrum.
There are no such restrictions at $\la=-\bar \la$, so the continuous
spectrum can be find in the imaginary axis.

Taking $U(x,y,\la)=ik(\la)\sz+Q(x,y,\la)$, where

$$
Q(x,y,\la)=\left (\begin{array} {cc}
\frac {i}{2}\frac {\la+1}{\la-1}(\cosh u-1)&(-\frac {i}{2}\frac {\la+1}{\la-1}
\sinh u+\frac {u_z}{2})e^{-2ix}\\
(\frac {i}{2}\frac {\la+1}{\la-1}\sinh u+\frac {u_z}{2})e^{2ix}&
-\frac {i}{2}\frac {\la+1}{\la-1}(\cosh u-1)\end{array} \right),
$$
and introducing the columns $\phi^{\pm}_1,\: \phi^{\pm}_2$ of the
matrix $\phi^{\pm},\:$ from (1.3a) we will obtain Volterra's
integrable equations of the direct problem:

$$
\phi^{-}_1(x,\la)=e_1+\int d\xi g^{-}_1(x-\xi,\la)
Q(\xi,\la)\phi^{-}_1(\xi,\la),\
\eqno(2.12)
$$
$$
\phi^{-}_2(x,\la)=e_2+\int d\xi g^{-}_2(x-\xi,\la)
Q(\xi,\la)\phi^{-}_2(\xi,\la).\
\eqno(2.13)
$$
In these equations $e_1=(1,0)^T,\:e_2=(0,1)^T$ and
$g^{-}_{1,2}(x,\la)$ are Green's "initial" functions:

$$
g^{-}_1(x,\la)=\theta(x) diag(1,\: e^{-2ik(\la)x}),\;\;\;
g^{-}_2(x,\la)=\theta(x) diag(e^{2ik(\la)x},\: 1).\eqno(2.14)
$$
From (2.12)-(2.13) it follows that

$$
\phi^{-}_1(x,\la)\in H(\Omega_1\cup \Omega_3),\;\;
\phi^{-}_2(x,\la)\in H(\Omega_2\cup \Omega_4),\eqno(2.15)
$$
where $H(\Omega_i)$ is the class of functions
($\Omega_i=\{\la:(i-1)\pi/2<\arg \la<i\pi/2\} ,i=1,...4$), that are
analytical in the domain $\Omega_i$.

Completely analogously we can show that
$$
\phi^{+}_1(x,\la)\in H(\Omega_2\cup \Omega_4),\;\;
\phi^{+}_2(x,\la)\in H(\Omega_1\cup \Omega_3).\eqno(2.16)
$$
From (2.12)-(2.13) we can easy obtain the integrable representations
for the scattering data:

$$
a(\la)=1+\int d\xi[Q_{11}(\xi,\la)\phi^{-}_{11}(\xi,\la)
+Q_{12}(\xi,\la)\phi^{-}_{21}
(\xi,\la)],\eqno(2.17)
$$
$$
b(\la)=\int d\xi[Q_{11}(\xi,\la)\phi^{-}_{12}(\xi,\la)
+Q_{12}(\xi,\la)\phi^{-}_{22}
(\xi,\la)]e^{-2ik(\la)\xi},\eqno(2.18)
$$
and hence
$$
a(\la)\in H(\Omega_1\cup \Omega_3).\eqno(2.19)
$$
From (2.18) and the similar expression for $c(\la)$ it follows that
generally speaking, $b(\la)$ and $c(\la)$ don't admit an analytical
extension from both the real axis and imaginary one. On assuming
$y=0$, in (2.18), accordingly to (2.10), we obtain the equality:

$$
\int d\xi[Q_{11}(\xi,0,\la)\phi^{-}_{12}(\xi,0,\la)+Q_{12}(\xi,0,\la)
\phi^{-}_{22}(\xi,0,\la)]e^{-2ik(\la)}=0 , \eqno(2.20)
$$
that connects the boundary values of the desired function and its
normal derivative and acts as the boundary condition (in term of the
spectral representation).

Let us introduce a matrix complex-valued function
$\Omega=\Omega(x,y)$, that can be defined as

$$
\Omega(x,y)=\lim \limits_{|\la| \to \infty} \Psi^{-}(x,y,\la) .   \eqno(2.21)
$$
From (2.21 ), (2.3) it follows, that the expression
$$
\Omega=e^{-ix\sz}e^{\frac {u}{2}\sx}\sy e^{-ix\sz}\bar \Omega\sy  \eqno(2.22)
$$
is fulfilled. This relation will be used later.

Matrix $\Omega$ obeys the equation:

$$
\Omega_x=U(x,y,\infty)\Omega,\eqno(2.23)
$$
and, in addition, $\det \Omega(x,y)=1.$ We suppose also that it has
asymptotics:
$$
\Omega_0=\lim \limits_{x \to +\infty} \Omega(x)=(-1)^Ne^{\frac{i\alpha_0}{2}\sz},
\eqno(2.24)
$$
$$
\lim \limits_{x \to -\infty}\Omega(x)=I, \eqno(2.25)
$$
where $\alpha_0$ is some parameter, $\alpha_0 \not\equiv 0$, and the
sense of the integer number $N$ will be clear further. The choice of
the asymptotics is in agreement with (1.3a), (2.23) and is stimulated
by the gauge equivalence of boundary problems.

From (2.2), (2.23), (2.24) we have:

$$
T(\infty)=(-1)^N\exp {(\frac {i\alpha_0}{2}\sz)}, \; \;\; T(0)=(-1)^N
\exp {(-\frac {i\alpha_0}{2}\sz)}, \eqno(2.26)
$$
and from (2.10) and the similar expression for $c(\la)$ it follows
that

$$
b(\pm 1-0)=c(\pm 1+0)=0.\eqno(2.27)
$$
\vskip0.5cm

\hfil{{\bf {3. Dispersion relations and identities of traces.}}\hfil
\vskip0.6cm

The results are obtained in the previous sections allows to write
down the dispersion relations and to obtain identities of traces as
well.

From (2.17) we have:

$$
a(\la)=(-1)^Ne^{\frac {i\alpha_0}{2}}[1+O(\frac {1}{|\la|})],\:\;
|\la|\to \infty,\:\; \la \in \Omega_1 \cup \Omega_3 .\eqno(3.1)
$$
Similarly it can be shown

$$
d(\la)=(-1)^Ne^{-\frac {i\alpha_0}{2}}[1+O(\frac {1}{|\la|})],\:\;
|\la|\to \infty,\:\; \la \in \Omega_2 \cup \Omega_4 .\eqno(3.2)
$$
We assume now that the coefficient $a(\la)$ has finite number of
simple zeros in domains $\Omega_1$ and $\Omega_3$, with the quantity
of the zeros being same and equal to $N$ in both domains. Let us also
assume  that the zeros don't belong to the continuous spectrum (may
be except of the zones accordingly to the formula (2.10)). Moreover,
we will consider that $a(\la)=a_s(\la)a_c(\la)$, where $a_s(\la)$ is
given by the contribution of the discrete part of the spectrum
("solitonic" part) and $a_c(\la)$ - to the continuous part spectrum.
Taking into account (2.2), (2.10)-(2.11) and equality
$|a_s(\la)|^2=1$ at $\la=\bar \la$, we will have

$$
\ln a_c(\la)+\ln d_c(\la)=0,\:\; \la=\bar \la,
$$
$$
\ln a_c(\la)+\ln d_c(\la)=\ln (1-|b(\la)|^2),\:\; \la=-\bar \la .
$$
These equalities assign the bound of piecewise analytical functions
$\ln a_c(\la)$ and $\ln d_c(\la)$ in the continuous spectrum. On
applying the Cochy's formula to the function $\ln a_c(\la) \in
H(\Omega_1
\cup
\Omega_3)$
and taking into account both (2.26) and two previous relations, we
obtain

$$
\ln a_c(\la)=-\int^{\infty}_{-\infty}\frac {d\mu}{2\pi i}
\frac {\ln(1-|b(i\mu)|^2)}{\mu+i\la}\:sign(\mu) .\eqno(3.3)
$$

Now let us consider the "soliton" part of the coefficients $a(\la)$:
$a_s(\la)$. Let $a_s(\la_n)=0$, where $\la_n \in
\Omega_1$. Then in a view of the involution $\bar a(\la_n)=a(-1/\bar
\la_n)e^{i\alpha_0}$ we have: $a(-1/\bar \la)=0,
\; -\bar {\la}^{-1} \in \Omega_3.$
From this and with consideration for (2.6) we will find the
expression for $a_s(\la)$:

$$
a_s(\la)=(-1)^Ne^{\frac {i\alpha_0}{2}}
\prod_{n=1}^N \frac {(\la-\la_n)(\la+\frac {1}{\bar \la_n})}
{\la+\bar \la_n)(\la-\frac {1}{\la_n})}\: .\eqno(3.4)
$$
Combining the equation (3.3) and (3.4),  we will finally obtain

$$
a(\la)= (-1)^Ne^{\frac {i\alpha_0}{2}}
\prod_{n=1}^N \frac {(\la-\la_n)(\la+\frac {1}{\bar \la_n})}
{\la+\bar \la_n)(\la-\frac {1}{\la_n})}\times
$$
$$
\times \exp {[-\int_{-\infty}^{\infty}
\frac {d\mu}{2\pi i}\frac {\ln (1-|b(i\mu)|^2)}{\mu+i\la}\:sign(\mu)]}.
\eqno(3.5)
$$
From the latter with consideration for (2.26) it follows:

$$
e^{-i\alpha_0}=\prod_{n=1}^N\frac {\la_n^2}{\bar \la_n^2}
\exp {[-\int_{-\infty}^{\infty}\frac {d\mu}{2\pi i}
\frac {\ln (1-|b(i\mu)|^2}{\mu}]} .
\eqno(3.6)
$$
In particular, at $b(\mu)=0$, i.e. in the "solitonic" sector of the
problem we have:

$$
e^{-i\alpha_0}=\prod_{n=1}^N\frac {\la_n^2}{\bar \la_n^2}\: .\eqno(3.7)
$$
It follows here from that at $\alpha_0=0$ and $N=1$ either
$\la_1=\bar
\la_1$ or $\la_1=-\bar \la_1$. In the first case the eigenfunction
proves to be non-localized and the solution doesn't fit to the
asymptotic (2.1). In the latter case ($\la_1=-\bar \la_1)\: \:\:
a_s(\la_1)=1$; means that discrete eigenvalue is in the continuous
spectrum (the situation that is beyond the consideration). Hence in
the reflectionless case at $\alpha_0=0$ the minimal value of $N$ is
equal two. This explains the cause of introducing the parameter
$\alpha_0$ in (2.24).

Assuming $\la_n=\rho_ne^{i\theta_n},\;\rho_n>0,
\; \theta_n \in (0,\pi/2)$, in (3.7),
we will obtain $\sum_{i=1}^N \theta_n=-\alpha_0/4+\pi n/2,\; n=0, \pm
1,
\pm 2,
\ldots $ ; in particular, at $N=1$:  $\theta_1=-\alpha_0/4+\pi k/2$, where
$k$ is the number to fulfill the inequalities $0 < -\alpha_0/4+\pi
k/2<\pi/2$.

Let $a_s(\la)=1$, i.e. discrete spectrum in the system is
unavailable. In this case from (3.6) we have:

$$
\alpha_0=-\int_0^{\infty}\frac {d\mu}{\pi}\frac {\ln (1-|b(i\mu)|^2)}
{\mu}\: .\eqno(3.8)
$$
From (3.5) and (3.8) it follow, that

$$
T(\pm 1)=1\:.\eqno(3.9)
$$
It should be noticed, that the expression of the type of (3.5) holds
for the coefficient $d(\la)$ also, however we will not discuss it
here.

Let us turn to deducing the identities of traces. Do to the matrix
$\Omega$ (the relation (2.23)) can't be determined explicitly the
equalities in the points zero and infinity can't be obtained (because
we don't know solutions of the Riccati's equation in this cases).
Therefore we will restrict ourselves by the consideration of the
relations near points of the poles divisors.

Assuming $\ln a(\la)=\sum_{p=0}^{\infty}a_p^{(\pm 1)}(\la \pm 1)^p,\;
\la \to \pm 1,$ from (3.5) we have: $a_p^{(1)}=-a_p^{(-1)}$, where $p$
is a even number, and $a_p^{(1)}=\bar a_p^{(-1)},$ where $p$ is an
odd number. These equalities being the consequence of the spectral
structure of the problem permit to control the validity of
intermediate calculations and the final result.

From (1.3a) using the standard way of calculations one can obtain
Riccati's equation for the function
$F(x,\la)=\phi^{-}_{21}(x,\la)/\phi^{-}_{11}(x,\la)$:

$$
F_x+Q_{12}F^2=(Q_{22}-Q_{11}-2ik(\la))F+Q_{21} .\eqno(3.10)
$$
Also from the equation for the function $\phi^{-}_1$, in view of
(2.2), it follows that

$$
\ln a(\la)= \int_{-\infty}^{\infty} dx[Q_{11}(x,\la)+Q_{12}(x,\la)F(x,\la)] .
\eqno(3.11)
$$
Using explicit form of the matrix elements $Q_{ij}$, at the vicinity
of the point $\la=-1$ we will obtain
$(F(x,\la)=\sum_{s=1}^{\infty}F_s^{(-1)}(\la+1)^s))$:
$F_1^{(-1)}=(iu_z/2)e^{2ix}$. Then accordingly to (3.11) $\ln
a(\la)=0,\:
\la \to -1$, which is in agreement with (3.9).
In the next order $(\sim (\la+1))$, considering (3.10), (3.11) and
(3.5), we will have:

$$
1+\frac {\rho_1^2-1}{\rho_1^2+2\rho_1\cos \theta_1+1}+
\int_{-\infty}^{\infty}\frac {d\mu}{2\pi}\frac {(\mu^2-1)\ln(1-|b(i\mu)|^2)}
{(\mu^2-1)^2+4\mu^2}\:sign(\mu)=
$$
$$
=\frac {1}{16}\int_{-\infty}^{\infty}dxu_x(x,0)u_y(x,0) , \eqno(3.12)
$$

$$
-\frac {\rho_1\sin \theta_1}{\rho_1^2+2\rho_1\cos \theta_1+1}+
\int_{-\infty}^{\infty}\frac {d\mu}{2\pi}\frac {\mu\ln(1-|b(i\mu)|^2)}
{(\mu^2-1)^2+4\mu^2}\:sign(\mu)=
$$
$$
=-\frac {1}{4}\int_{-\infty}^{\infty} dx [(\cosh u(x,0)-1)-\frac {1}{8}
(u_x^2(x,0)-u_y^2(x,0))] , \eqno(3.13)
$$
where $\la_1=\rho_1e^{i\theta_1},\;\theta_1 \in (0,\pi/2).$ Thus, the
continuous spectrum of the problem contributes the left parts of the
identities of traces. It should be noted that the similar character
the identities have also in the neighbourhood of the point $\la=1$.

\vskip0.6cm

\hfil{{\bf {4. Equations of the inverse problem and formulas for reconstruction.}}
\hfil
\vskip0.6cm

It follows from the analysis of properties of the operator spectrum
of an associated linear problem and analytical properties of Jost's
solutions performed above that the continuous spectrum of the problem
is in "the cross": $\{\la: \la_R=0 \cup \la_I=0\}$, whereas matrix
columns of these solutions admit the analytical extension to certain
domains of the plane of a complex spectral parameter $\la$. This
allows using Riemann's technique problem. However there exists some
additional difficulty here: from (2.21) and the determination of the
matrix $\phi^{-}$ it follows, that
$\phi^{-}(x,\infty)=\Psi(x,\infty)=\Omega(x,y)$, with the matrix
$\Omega(x,y)$ being unknown. This leads to Riemann's local problem on
an analytical factorization appears to be characterized by
noncanonical conditions at the infinity in this case. Also some
difficulties arise in searching for formulas of the reconstruction.
This section is devoted to these problems.

As the method of solution of the inverse problem we choose the method
of Riemann's vector problem. From (2.2) and the connection
$\Psi^{\pm}$ and $\phi^{\pm}$ we obtain:

$$
\phi^{-}(x,\la)=\phi^{+}(x,\la)T_1(x,\la)\: ,\eqno(4.1)
$$
where

$$
T_1(x,\la)=\exp (ik(\la)x\sz)T(\la)\exp(-ik(\la)x\sz).
$$
We use one of the equalities following from (4.1)

$$
\frac {\phi^{+}_1(x,\la)}{d(\la)}=\phi^{-}_1(x,\la)-\phi^{-}_2 (x,\la)
\frac {c(\la)}{d(\la)}e^{-2ik(\la)x}\: .\eqno(4.2)
$$
Now let us suppose that $\la_n \in \Omega_1$ is such that
$a(\la_n)=0$ and also $-1/\bar \la_n  \in \Omega_3$ and $a(-1/\bar
\la_n)=0$. Then the vector-functions
$\Psi^{-}_1(x,\la)$ and $\Psi^{+}_2(x,\la)$ proved to be linear-
dependent at these points, from here we have:

$$
\phi^{-}_1(x,\la_n)=-\frac {1}{b_n}\phi^{+}_2(x,\la_n)e^{-2ik(\la_n)x} ,
\eqno(4.3)
$$
$$
\phi^{-}_1(x,-\frac {1}{\bar \la_n})=
\frac {1}{\bar b_n}\phi^{+}_2(x,-\frac {1}{\bar \la_n})
e^{-2ik(-\frac {1}{\bar \la_n})x} ,\
\eqno(4.4)
$$
where $\phi^{-}_1(\la_n),\:\phi^{+}_2(\la_n),\:\phi_1^{-}(-\bar \la^{-1}_n),\:
\phi^{+}_2(-\bar \la_n^{-1})$ are eigenfunctions of the discrete spectrum,
$b_n$ are the coefficients of transition.

Taking into account that $\phi^{-}_1(\la) \to \Omega^1_1(x,y),\;
|\la|
\to
\infty,\;
\la \in \Omega_1 \cup \Omega_3$, and $\phi^{+}_1(\la)/d(\la) \to \Omega^1_1(x,y),\:
\la \in \Omega_2 \cup \Omega_4$, where $\Omega^1_1=\Omega(x,y)e_1$,
and applying Cochy's formulas to the vector-column
$(\phi^{-}_1-\Omega^1_1)$ that is analytical in $\Omega_1$ and also
using (4.2)-(4.4), we obtain the following equality:

$$
\phi^{-}_1(\la)=\Omega^1_1-i\sum_{n=1}^N\frac {\beta_n}{\la^2_n(\la-\frac
{1}{\la_n})}e^{-2ik(\frac {1}{\la_n})x}\chi (x)
\bar \phi^{-}_1(-\frac {1}{\bar \la_n})+
$$
$$
+i\sum_{n=1}^N\frac {\bar \beta_n}{\la+\bar \la_n}\chi(x)
\bar \phi^{-}_1(\la_n)+
\int_{-\infty}^{\infty}\frac {d\mu}{2\pi}\frac {\bar b(\mu)}{\bar a(\mu)}
\frac {\chi(x)}{\mu+\la}e^{-2ik(\mu)x}\bar \phi^{-}_1(\mu)sign(\mu)-
$$
$$
-\int_{-\infty}^{\infty}\frac {d\mu}{2\pi}\frac {\bar b(i\mu)}{\bar a(i\mu)}
\frac {\chi(x)}{\mu+i\la}e^{-2ik(i\mu)x}\bar \phi^{-}_1(i\mu)sign(\mu),\
\eqno (4.5)
$$
where $\chi(x)=e^{-ix\sz}\sy e^{-iu/2\sx}e^{-ix\sz},\; \beta_n(y)=
b_n/a^{\prime}(\la_n)=\beta_n(0)e^{2l(\la_n)y}.$ The equation is not
closed equation for the column $\phi^{-}_1(\la)$ yet due to it
involves an unknown vector-function $\Omega^1_1$. Therefore to obtain
the necessary equation we will use (2.22) rewritten in the form:

$$
\Omega^{-1}e^{-ix\sz}\sy e^{-\frac {u}{2}\sx}e^{-x\sz}\bar \Omega=\sy \: .
\eqno(4.6)
$$
Substituting (4.6) into (4.5) and putting

$$
\phi^{-}_1(\la)=\Omega(x,y)\tilde \phi^{-}_1(\la) ,\eqno(4.7)
$$
we obtain the following system of singular integral equations for the
function $\tilde \phi^{-}_1(\la)$:

$$
\tilde \phi^{-}_1(\la)=e_1+
i\sum_{n=1}^N\frac {\bar \beta_n
e^{-2ik(\bar \la_n)x}}{\la+\bar \la_n}\sy\bar {\tilde \phi^{-}_1}(\bar \la_n)
$$
$$
-i\sum_{n=1}^N\frac {\beta_ne^{-2ik(\frac {1}{\la_n})x}}{\la_n^2(\la-
\frac {1}{\la_n})}\sy\bar {\tilde \phi^{-}_1}(-\frac {1}{\bar \la_n})+
\int_{-\infty}^{\infty}\frac {d\mu}{2\pi}\frac {\bar b(\mu)}{\bar a(\mu)}
\frac {e^{-2ik(\mu)x}}{\mu+\la}\sy\bar {\tilde {\phi^{-}_1}}(\mu)sign(\mu)-
$$
$$
-\int_{-\infty}^{\infty}\frac {d\mu}{2\pi}\frac {\bar b(i\mu)}{\bar a(i\mu)}
\frac {e^{-2ik(i\mu)x}}{\mu+i\la}\sy\bar {\tilde \phi^{-}_1}(i\mu)sign(\mu),\
\eqno(4.8)
$$

$$
\tilde \phi^{-}_1(\la_p)=e_1+
i\sum_{n=1}^N\frac {\bar \beta_n
e^{-2ik(\bar \la_n)x}}{\la_p+\bar \la_n}\sy\bar {\tilde \phi^{-}_1}
(\bar \la_n)-
$$
$$
-i\sum_{n=1}^N\frac {\beta_ne^{-2ik(\frac {1}{\la_n})x}}{\la_n^2(\la_p-
\frac {1}{\la_n})}\sy\bar {\tilde \phi^{-}_1}(-\frac {1}{\bar \la_n})+
\int_{-\infty}^{\infty}\frac {d\mu}{2\pi}\frac {\bar b(\mu)}{\bar a(\mu)}
\frac {e^{-2ik(\mu)x}}{\mu+\la_p}\sy\bar {\tilde {\phi^{-}_1}}(\mu)sign(\mu)-
$$
$$
-\int_{-\infty}^{\infty}\frac {d\mu}{2\pi}\frac {\bar b(i\mu)}{\bar a(i\mu)}
\frac {e^{-2ik(i\mu)x}}{\mu+i\la_p}\sy\tilde {\bar \phi^{-}_1}
(i\mu)sign(\mu),\
\eqno (4.9)
$$
$$
\tilde \phi^{-}_1(-\frac {1}{\bar {\la_p}})=e_1+
i\sum_{n=1}^N\frac {\bar \beta_n
e^{-2ik(\bar \la_n)x}}{\bar \la_n-1/\bar \la_p}\sy\bar {\tilde \phi^{-}_1}
(\bar \la_n)+
$$
$$
+i\sum_{n=1}^N\frac {\beta_ne^{-2ik(\frac {1}{\la_n})x}}
{\la_n^2(\frac {1}{\bar \la_p}+
\frac {1}{\la_n})}\sy\bar {\tilde \phi^{-}_1}(-\frac {1}{\bar \la_n})+
\int_{-\infty}^{\infty}\frac {d\mu}{2\pi}\frac {\bar b(\mu)}{\bar a(\mu)}
\frac {e^{-2ik(\mu)x}}{\mu-\frac {1}{\bar {\la_p}}}
\sy\bar {\tilde {\phi^{-}_1}}(\mu)sign(\mu)-
$$
$$
-\int_{-\infty}^{\infty}\frac {d\mu}{2\pi}\frac {\bar b(i\mu)}{\bar a(i\mu)}
\frac {e^{-2ik(i\mu)x}}{\mu-i\frac {1}{\bar {\la_p}}}\sy
\bar {\tilde \phi^{-}_1}
(i\mu)sign(\mu).\
\eqno (4.10)
$$
The equations (4.9), (4.10) are obtained from (4.8) at $\la=\la_p$
and $\la=-1/\bar \la_p$ respectively. On the right-hand sides of
these equations it is necessary to take into account the availability
of the zones (2.10).

The system (4.8)-(4.10) is the system of the equations for the
eigenfunctions of the continuous and discrete spectrum. Considering
the system is written for the "gauge" vector-functions, the necessary
formulas of the reconstruction should be written in terms of the same
function.

Using (1.3a), (4.7), we obtain the differential equation for the matrix
$\tilde \phi^{-} \equiv \tilde \phi^{-}(x,\la)$:

$$
\tilde \phi^{-}_x=-ik(\la)\tilde \phi^{-}\sz+\frac {1}{\la-1}
\Omega^{-1}[i(\cosh u-1)\sz+\sinh ue^{-2ix\sz}\sy]\Omega\tilde \phi^{-}+
$$
$$
+ik(\la)\Omega^{-1}\sz\Omega\tilde \phi^{-} . \eqno(4.11)
$$
In the neighborhoods of the point $\la=1$,
setting $\tilde \phi^{-}=\sum_{k=0}^{k=\infty}
\tilde \phi_k(\la-1)^k  \; (\tilde \phi^{-}_0 \to I, \;x \to -\infty),$
we find:

$$
\tilde \phi^{-}_{0}=\Omega^{-1}\left(\begin{array}{cc}
1  &  \tanh \frac {u}{2}e^{-2ix}\\
\tanh \frac {u}{2}e^{2ix}& 1\\ \end{array}\right) .\eqno(4.12)
$$

Now we parameterize the unknown matrix $\Omega$ by four complex
functions of variables $x,\:y$:

$$
\Omega (x,y)=\left(\begin{array}{cc}
\alpha(x,y)&\beta(x,y)\\
\gamma(x,y)&\delta(x,y)\\ \end{array}\right) .\eqno(4.13)
$$
Then the relation (4.6) gives 4 algebraic connections:

$$
\alpha=\bar \delta \cosh \frac {u}{2}-\bar \beta \sinh \frac {u}{2}e^{-2ix}, \
\;  \beta=\bar \alpha \sinh \frac {u}{2}e^{-2ix}-\bar \gamma \cosh \frac {u}{2},
$$
$$
\gamma=-\bar \beta \sinh \frac {u}{2}+\bar \delta \cosh \frac {u}{2}e^{2ix} ,\ \;
\delta=\bar \alpha \cosh \frac {u}{2}-\bar \gamma \sinh \frac {u}{2}e^{2ix} .
$$
In a view of the relation and from (4.12) we obtain ($\tilde
\phi^{0}=
\tilde \phi_{0}$):
$$
\tilde \phi^{0}=\frac {1}{\cosh \frac {u}{2}}\bar \Omega^{T} , \eqno(4.14)
$$
where $\det \tilde \phi^{0}=(\cosh \frac {u}{2})^{-2}$. Combining
(4.12), (4.14) and (4.6), we will have:

$$
\cosh \frac {u(x,y)}{2}=[|\tilde \phi^{0}_{11}|^2+|\tilde \phi^{0}_{21}|^2]^{-1}
=[|\tilde \phi^{0}_{12}|^2+|\tilde \phi^{0}_{22}|^2]^{-1} , \eqno
(4.15)
$$
or
$$
\cosh \frac {u(x,y)}{2}=[<\tilde {\phi^{0}_1}^T,\tilde \phi^{0}_1>_2]^{-1}=
[<\tilde {\phi^{0}_2}^T,\tilde \phi^{0}_2>_2]^{-1} ,\eqno(4.16)
$$
where the symbol $< , >_2$ denotes the scalar product of the vectors
in $\mathbb C^2$. Another equivalent representation of (4.16) is

$$
\cosh \frac {u(x,y)}{2}=\{\frac {1}{2} Tr(\tilde {\phi^{0}}^T
{\bar {\tilde \phi^0})}\}^{-1}  . \eqno(4.17)
$$

Analogously we can obtain one more formulas of the reconstruction:

$$
\sinh \frac {u(x,y)}{2}=\frac {2\tilde Ae^{-2ix}}{1+\sqrt {1-4\tilde A^2
e^{-4ix}}}\: ,\eqno(4.18)
$$
where  $\tilde A=<(((\phi^{-})^0)_1)^T,(((\phi^{-}))^0)_2>_2$,
$\tilde A \to 0$ at $|x| \to \infty$, and the requirement of reality
of the potential leads to the value $\tilde Ae^{-2ix}$ is to be real.
It is easy to prove the latter using (4.14) and the definition of
$\tilde A$. Furthermore, employing elementary transformations it is
demonstrated that (4.18) is equivalent to (4.16).

From (4.15), (4.18) it follows that the solution of the problem
(1.1), (1.2) has a nonsingular character.

In solitonic sector of the model ($b(\la)=0)$ one can be found the
expression for the formulas of the reconstruction through
eigenfunction of the discreet spectrum. can find the expression in
terms of the eigenfunctions of discrete spectrum. For this purpose we
first resolve the right of (4.8) at $\lambda \to 1$ and after
equating it to ${\tilde \Phi}_1^0$ we will obtain from (4.15) the
following:

$$
u(x,y)=2\ln \frac {1-\sqrt {1-(|\xi|^2+|\eta|^2)^2}}{|\xi|^2+|\eta|^2} ,
\eqno(4.19)
$$
where
$$
\xi=\xi(x,y)=\sum_{n=1}^{2N}\frac {\bar {\beta_n}e^{-2ik(\bar \la_n)x}}
{\bar \la_n+1}\bar {\tilde \phi^{-}}_{11}(\la_n) ,
$$
$$
\eta=\eta(x,y)=1+\sum_{n=1}^{2N}\frac {\bar {\beta_n}e^{-2ik(\bar \la_n)x}}
{\bar \la_n+1}\bar {\tilde \phi^{-}}_{21}(\la_n) ,
$$
$$
\la_{n+N}=-\frac {1}{\bar {\la_n}}, \; \beta_{n+N}=-\frac {\bar \beta_n}
{\bar {\la_n}^2}, \;\; n=1,2,...  N.
$$
The functions $\tilde \phi^{-}_{11}(x,\la_n),\;\tilde
\phi^{-}_{21}(x,\la_n)$,
involved in (4.19) can be found from the system of linear algebraic
equations (4.9)-(4.11). Then:

$$
\xi=<\bar {\Lambda} \bar E,\; R^{-1}E>_{2n},\;\;
\eta=1-<\bar \Lambda \bar E, R^{-1}A\bar E>_{2n},\eqno(4.20)
$$
where:

$A=\bigl \{A_{mn}\bigr \},\:\:A_{mn}=E_m{\bar E}_n/(\la_m+\bar \la_n),\:\:
E_m=\beta^{1/2}_m e^{ik(\la_m)x}$,

$\Lambda=diag{((\la_1+1)^{-1},
....(\la_{2n}+1)^{-1})},\:\:m,\:n=1,....2N,\:\:E=(E_1,....E_{2N}),$

$\:R=I+A{\bar A}$.

N-"soliton" solution (4.19), (4.20) is the exact solution of the
initial boundary problem. However in the simplest case ($N=1$) it is
it is rather cumbersome and it is beyond the paper.

Also it should be noted that the asymptotics of the solution for the
continuous spectrum has been calculated in [6] proceeding from gauge
equivalence between the given problem and the problem for
Heisenberg's ferromagnet, with the technique of the ref.[7] being
used essentially.

\vskip0.6cm

\centerline{{\bf {5. Application to the model of one-charged plasma}}}
\hfil{\bf at negative temperatures.}
\vskip0.6cm

This model is lead by the model of the equilibrium system of charged
particle (vortices) on the plane [2, 3]. The latter permits
generalization of Debay-Hukkel classical model [8] for plasma or an
electrolyte solution to the case of a nontrivial statistical "phone".

Following [2], [3], let us suppose that there is the Coulomb's system
of $N^{+}_0$ particles possessing a charge $+e$ and $N^{-}_0$
particles possessing a charge$-e$ embedded in some finite volume
$V=L^2$ (where $L$ is the linear scale). Furthermore, we will suppose
that a temperatures of the ionic component coincides with a
temperature of electronic one, with both of them being same and equal
to $T$ (Boltsmann's constant is considered to be equal to unit).
Using the BBGKI chain it is convenient to consider a scalar potential
$\Phi=\Phi(x,y)$:

$$ e\Phi(x,y)=n_0\int dx^{\prime}dy^{\prime} \phi({\bf r}-{\bf r}^
{\prime})\rho({\bf r^{\prime}}), \;\; {\bf r}=(x,y) , \eqno(5.1)
$$
where $n_0$ is the average density of charge particles, $\phi({\bf
r}-{\bf r}^ {\prime})$ is Coulomb's two-particles interaction,
$\rho(\bf r)$ is a difference between the ionic density and an
electron one. Then employing rather rigorous procedure of the
reduction of the chain we can obtain the equation:

$$ \triangle \Phi=\frac {4\pi n_0c}{l^0}[\exp(\frac
{e\Phi}{T}) -\exp(-\frac {e\Phi}{T})], \eqno(5.2) $$ where $l^0$ is a
characteristic length,

$$ c=\frac {V}{\int dxdy[\exp[\frac
{n_0}{T}\int dx^{\prime} dy^{\prime} \phi({\bf r}-{\bf r}^{\prime})
\rho(\bf r^{\prime})]]}\:.
$$
Setting $u=\frac {e\Phi}{T}$ and Debay's radius $\tilde
\triangle=(4\la_D^2/c)\triangle ,\; \la_D^2=l^0|T|/(8\pi
n_0e^2)$ in (5.2), we find:

$$ sign(T)\tilde {\triangle}
u=4\sinh u .\eqno(5.3) $$

This equation leads to the equation (1.1) at $T < 0$ ( the case $T
> 0$ is considered in [9, 10]).

According to the statement of the problem describe above the equation
(5.3) held for $\mathbb R^2_{+}$ should be added by the given value
of the field at the boundary ${\partial \mathbb R^2_{+}}$
considering.

Now we consider some physical aspects of the results obtained in the
previous sections. Assuming $n_0=N^{\pm}_0/V=const$, i.e in the
thermodynamical limit.

For this purpose we will reveal some features of the behaviour of the
equation (4.19). The total energy of the finite system of discrete
isolated charges for the two-dimensional case [2], [3] is:

$$
E_{tot}=\frac {1}{2}\sum_{i,j}(N^{+}_{{0}_i}-N^{-}_{{0}_i})\Phi_{ij}
(N^{+}_{{0}_j}-N^{-}_{{0}_j}) , \eqno(5.4)
$$
where $N^{+}_0=\sum_{i}N^{+}_{{0}_i}, \; N^{-}_0=\sum_{j}N^{-}_{{0}_j},\:
\Phi_{ij}=-(2e^2/l^0)\ln |\bf r_i-\bf r_j|$ is Coulomb's potential of
the interaction between the particles of $"i"$ and $"j"$ clusters,
${\bf r_i}$ and ${\bf r_j}$ are their radius-vectors.

Comparing the expressions for $\Phi_{ij}$ with (4.19), we can see
they have the similar form. Let $B=|\xi|^2+|\eta|^2$ ,   and $0
\le B \le 1$, due to the potential is real. Then from (4.19) it follows,
 that $u(x,y)\simeq
2\ln (B/2)
\to -\infty$ at $B
\to 0$ and $u(x,y) \to 0$ at $B \to 1$. Here in the terms of eigenfunctions of
the discrete spectrum and values $\xi,\:\eta$, the first condition
means that: $\xi,\:\eta \to 0$, which leads to $\sum_{n=1}^{2N} [\bar
\beta_n\exp(-2ik(\bar \la_n)x)/(\bar
\la_n+1)]\tilde \phi^{-}_{21}( \la_n) \to -1$; it is possible on the case
of finite ${\bf r}$. The latter can be fulfilled, in particular, at
$\xi \to 0,\;\eta \to 1$, i.e. at $r \to \infty$.

Thus, we have been obtained the following qualitative picture. Let us
suppose there is no external field first ($u(x,0)=u_y(x,0)=0$). Also
clusters of the same charges can by formed from the charged particle.
After appearance of an external field the space distribution being
available before changes into another stationary state. At the any
point $(x,y)
\in \mathbb R^2_{+}$ the potential is the sum of potential arisen
by the the boundary $\partial
\mathbb R^2_{+}$ and potential of system; as result, some
self-consistent field is created. The potential of the field is
described by expression $Tu(x,y)/e$, where $u(x,y)=u(B)$, and the
function $B=B(x,y)$ plays the role of the "transmission" function of
the medium. At certain finite $x,\:y$ a local and deep minimum of the
potential corresponds to the minimal value of $B$, with $u < 0$ at
the vicinity of the minimum. Considering $T < 0$ a value $\Phi$
proves to to be big and positive. This fact shows a local
concentration of particles of the identical sign. Due to non-singular
character of (4.19) $\Phi$ keep be bounded up and it leads to a
limitation of the energetic spectrum of the system. So we deal with
an abnormal system where the inversion of the level distribution is
available, it means that the positive value of the energy together
with negative temperatures leads to the higher levels (in average)
become more preferred for the system than the lower ones.
Furthermore, the charges distribution of clusters with same sign
appears more apparently.

The condition $r \to \infty$ and $u(B) \to 0$ corresponds to the
maximum of $B$, i.e. as we recede from the boundary the field created
by those charges fads due to the interaction with own field of the
medium. The damping is slower than in the similar situation when
temperatures are positive [9,10]. This is not difficult to see on
comparing the solution of corresponding linear problems:

$$
\triangle u=\pm 4u,\;\; u \to 0 \;\;\; at\;\;\; r \to \infty , \eqno(5.5),
$$
where the sign "+" corresponds to system with $T > 0$, whereas "-"
corresponds to negative temperatures. In the former case Green's
function is proportional to McDonald's function of the zero index and
has the asymptotic: $\sim (e^{-x}/\sqrt x) [1+O(1/x)], \; x
\to
\infty $, and in the latter case it is proportional to Neemann's function of zero
index that behaves like: $(\sin (x-\pi/4)/\sqrt x)[1+O(1/x)],\; x
\to
\infty.$ Therefore the slower damping of the field at $T < 0$ seems to be explained by
weaker defense of the boundary due to an alternation of domains with
the same charged particles.

The picture describe above is true everywhere on the semi-plane
except of the certain directions defined evidently. In these cases
$B$ takes an intermediate values: $B
\in [\epsilon, 1-\epsilon],
\;\epsilon > 0$, and the potential is an oscillating
function in those directions, i.e. there is the distant order in the
medium. It means that a coherent structure of the distribution of an
electric field and densities of a plasma distribution arises.

The author is grateful to V.D.Lipovski for useful discussions.

This work has been due to a financial support of The Russian
Foundation for Fundamental Research (Projects 98-01-01063,
00-01-00480).

\vskip1cm

\hfil{\bf {References}}

\vskip0.8cm
[1]. A.I.Bobenko, Uspechi Math. Nauk, {\bf 46}, ü4(1991)(in Russian).
\vskip0.3cm

[2]. G.Goyce and D.Montgomeri, J.Plasma Phys., {\bf 10}, ü1, 10(1973).
\vskip0.3cm

[3]. D.Montgomeri and G.Goyce, Phys.Fluids, {\bf 17}, ü6, 1139(1974).
\vskip0.3cm
[4]. Yu.B.Rumer and M.Ch.Rivkin. Thermodynamics, statistical physics and

kinetic. Moscow, Nauka (1977)(in Russian).

\vskip0.3cm
[5]. E.Sh.Gutshabash, Vestnik LGU, ser.Fisika, Chimia , {\bf 4}, 84(1990)(in Russian).

\vskip0.3cm
[6]. G.G. Varzugin, E.Sh.Gutshabash and V.D.Lipovski, Teoreticheskaya i

Matematicheskaya Fizika, {\bf 104}, ü3, 513(1995)( in Russian).

\vskip0.3cm
[7]. P.Deift and X.Zhou, Ann.of Math., {\bf 137}, ü2, 295(1993).

\vskip0.3cm

[8]. L.D.Landau and E.M.Lifchits. Statistical Physics. Moscow, Nauka,1982

(in Russian).

\vskip0.3cm
[9]. V.D.Lipovski and S.S.Nikulichev, Vestnik LGU, ser.Phisica, Chimia

{\bf 4}, 61(1989)(in Russian).

\vskip0.3cm
[10]. E.Sh.Gutshabash, V.D.Lipovski and S.S. Nikulichev, Teoreticheskaya

i Matematicheskaya Fizika, {\bf 115}, 323(1998)(in Russian); nlin.SI/0001012.

\end{document}